%% file: sample-sigplan.tex
\documentclass[sigplan,screen]{acmart}
\usepackage{soul}
\renewcommand\footnotetextcopyrightpermission[1]{} 
\AtBeginDocument{%
  \providecommand\BibTeX{{%
    \normalfont B\kern-0.5em{\scshape i\kern-0.25em b}\kern-0.8em\TeX}}}

\setcopyright{acmcopyright}
\copyrightyear{2023}
\acmYear{2023}
\acmDOI{XXXXXXX.XXXXXXX}

\acmConference[HASP '23]{Hardware and Architectural Support for Security and Privacy }{October 28--November 01,
  2023}{Toronto, Canada}
\acmSubmissionID{5}

\usepackage{multirow}
\usepackage{tabularray}
\begin{document}

\title[]{A First Order Survey of Quantum Supply Dynamics and Threat Landscapes}


\author{Subrata Das}
\email{sjd6366@psu.edu}
\affiliation{%
  \institution{Pennsylvania State University}
  \city{State College}
  \state{PA}
  \country{USA}
}

\author{Avimita Chatterjee}
\email{amc8313@psu.edu}
\affiliation{%
  \institution{Pennsylvania State University}
  \city{State College}
  \state{PA}
  \country{USA}
}

\author{Swaroop Ghosh}
\email{szg212@psu.edu}
\affiliation{%
  \institution{Pennsylvania State University}
  \city{State College}
  \state{PA}
  \country{USA}
}







\begin{abstract}
Quantum computing, with its transformative computational potential, is gaining prominence in the technological landscape.
As a new and exotic technology, quantum computers involve innumerable Intellectual Property (IP) in the form of fabrication recipe, control electronics and software techniques, to name a few. Furthermore, complexity of quantum systems necessitates extensive involvement of third party tools, equipment and services which could risk the IPs and the Quality of Service and enable other attack surfaces. This paper is a first attempt to explore the quantum computing ecosystem, from the fabrication of quantum processors to the development of specialized software tools and hardware components, from a security perspective. By investigating the publicly disclosed information from industry front runners like IBM, Google, Honeywell and more, we piece together various components of quantum computing supply chain. 
We also uncover some potential vulnerabilities and attack models and suggest defenses. We highlight the need to scrutinize the quantum computing supply chain further through the lens of security. 
\end{abstract}



\keywords{Quantum Computing, Quantum Supply Chain,
Intellectual Property, Security Threats, Vulnerabilities}



\maketitle
\pagestyle{plain}
\section{Introduction}
\label{sec:intro}

Quantum computing has the potential to revolutionize fields from cryptography to material science \cite{nielsen2001quantum, rol2020control}. But beneath the allure lies a complex infrastructure that supports the development, deployment, and operation of quantum computers. This infrastructure, from the fabrication of quantum bits (qubits) and cryogenic control units to the ultra-specific software that governs them, is a vast web of interconnected systems, operations, and technologies \cite{alberts2021accelerating}. 
Compared to classical computers, quantum computers are far more complex and require a very close collaboration among material scientists, physicists, engineers from various disciplines and computer scientists, to name a few, to make the system work. These systems are fragile and, as such, require frequent repair, servicing and calibration to maintain their presence in the market. Compared to a classical cloud computing environment where the faulty units can be disabled without a noticeable impact on quality of service (QoS), repair of quantum computers may disrupt the QoS significantly. Therefore, the reliability of the individual components of a quantum computer (besides the quantum processor) is key to maintaining QoS, profit and scientific endeavors. Many companies (who may be leaders in one aspect of quantum computing but not in other areas) are in the race to commercialize their qubit technologies. As such, they are forced to rely on tools and services from third parties not only to reduce the cost but also to bridge the knowledge gap. This situation is very similar to classical integrated circuit (IC) design that relies on third parties for software tools and Intellectual Property (IP) blocks but much exacerbated in the quantum domain.  

The supply chain for classical ICs starts with integrating external IP designs with internal designs, validating them and creating the IC layout, often in a GDS-II format \cite{rostami2014primer}. Internal or third party tools are used in the design and validation process. The final layout is sent to a foundry, where a mask is developed, and the ICs are manufactured. The chips are then tested at the manufacturing site and possibly at third-party test facilities. The fault-free chips are packaged, tested and sold. This entire supply chain is spread across multiple countries and hence, can raise security concerns. Therefore, the industry follows a standard procedure for securing the supply chain of classical ICs, which includes measures such as designing with security in mind, collaborating with trusted foundries, rigorous testing, implementing cybersecurity practices, and maintaining transparency.

\begin{figure}[h]
  \centering
  \includegraphics[width=\linewidth]{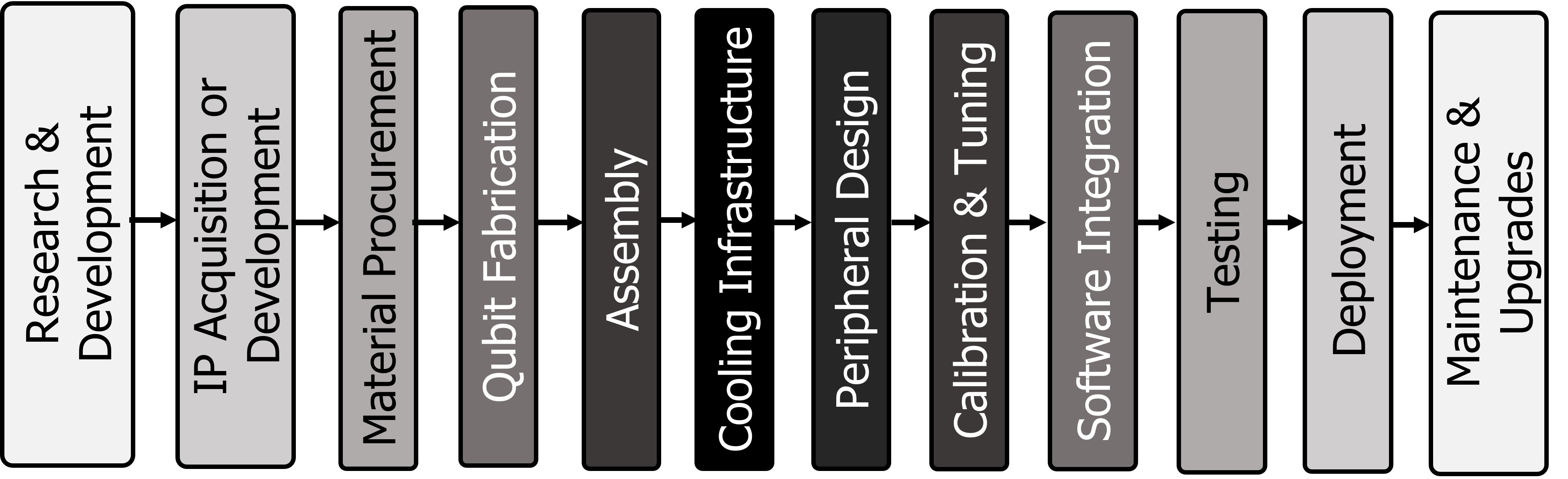}
  \caption{Supply chain of a superconducting quantum computer.}
  \label{fig:supplychain}
\end{figure}

Although the supply chain of conventional semiconductor technologies has matured over several decades, providing a robust and well-understood foundation, the supply chain for quantum technologies is still in its formative stages. This is primarily due to the fact that classical processors are manufactured in high volume whereas quantum processors are produced at extremely small scale.  
While specifics might vary between quantum computing architectures (e.g., superconducting qubits \cite{devoret2013superconducting} versus trapped ions \cite{haffner2008quantum}), a general supply chain might look something like Figure \ref{fig:supplychain}. Initially, extensive research is conducted to determine the optimal design and technology. This involves collaboration between scientists and engineers to ensure precision. Companies then choose to either design their own quantum blueprints or acquire pre-existing ones. Specialized materials, of exceptional purity, are procured for the components. Qubits are then fabricated in clean rooms. Following this, quantum processor is integrated with components like control systems and other critical parts to form the quantum computer. These computers require a very low-temperature environment for optimal functionality, necessitating advanced cooling systems. Once assembled, the system undergoes calibration to ensure its components interact correctly. Software is then incorporated, allowing users to interface with and utilize the quantum capabilities. Rigorous testing is carried out to ascertain its readiness. 
Quantum computers are 
often accessed remotely via online platforms. Owing to the nascent nature of quantum technology, periodic maintenance and upgrades are essential.

Furthermore, quantum computing systems may include multitude of IPs embedded in various layers of the stack \cite{trochatos2023hardware, ghosh2023primer}. For example, the real coupling map of the superconducting qubit processor (before disabling the non-functional links), pulsing techniques and frequencies could be an IP. Similarly, the materials used for fabricating the qubits could be an IP. The techniques to suppress crosstalk and noise could also be an IP. At the software level, the techniques to optimize the gate count and circuit depth could represent an IP.

There are known methodologies to protect IPs and establish trust even using untrusted components and parties in classical ICs whereas such methods are not known in the quantum computing domain. The supply chain of quantum computers, software stack, tools and associated services is still evolving. Various companies follow their own supply chain models at their convenience. Companies often employ a mix of in-house facilities for IPs, personalized hardware, or different vendors for components such as dilution refrigerators. Most commonly, companies conduct assembly and testing independently. This lack of standardized clarity within the quantum supply chain exacerbates security risks \cite{ghosh2023primer, xu2023exploration}. 

\vspace{-12pt}



\textbf{Contributions:}
We review the design of a representative quantum computing system including hardware, software, firmware, peripherals, and services (Section 2). This is followed by a description of potential IPs embedded in various layers of quantum stack (Section 3). 
We provide a comprehensive study of the quantum supply chain for prominent companies in the USA synergistic with a recent NSF-OSTP workshop on cybersecurity of quantum computing \cite{Cyber_workshop} (Section 4). 
We also highlight possible vulnerabilities in different layers of the quantum supply chain, attack models and defenses (Section 5). While investigating quantum-specific threats, we underscore that quantum technologies nonetheless inherit many conventional hardware and software vulnerabilities.

\section{General Architecture of a Quantum Computer}
\label{sec:architecture}

\begin{figure}[h]
  \centering
  \includegraphics[width=0.7\linewidth]{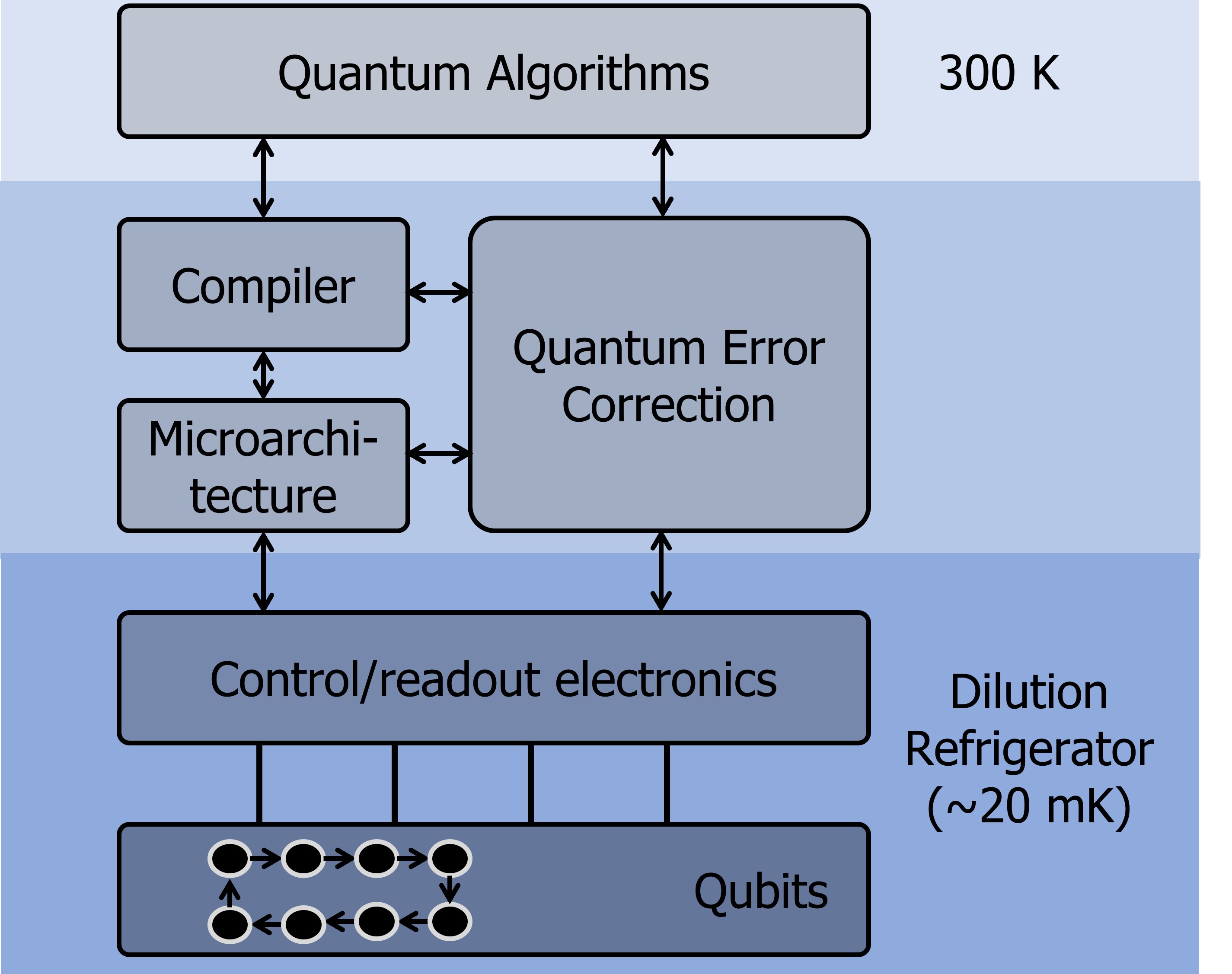}
  \caption{General architecture of a quantum computer. Three blue boxes with color gradients indicate the varying temperature required for different components of the system.
  }
  \label{fig:architecture}
  \vspace{-0.3cm}
\end{figure}

The architecture of a quantum computer spans from the high-level description of quantum algorithms to the actual physical operation of the qubits. This section will provide an overview of this general architecture (Figure \ref{fig:architecture})  using superconducting quantum computers as an illustrative example. 

\subsection{Quantum Algorithms and Compilers }

At the highest level of the quantum computer sits the quantum algorithms, often expressed in a high-level programming language like Q\# or Quil. These algorithms operate on idealized logical qubits and gates, without errors \cite{cross2017open}. 
The compiler translates the logical operations into instruction sequences that manipulate the physical qubits to implement the desired algorithms. 
The compiler schedules and optimizes these circuits while accounting for constraints imposed by qubit connectivity, gate fidelities etc. The pulse sequences and instructions generated by the compiler are then passed to the control electronics, which translate them into analog signals sent to the qubits at precisely timed intervals. Error mitigation and/or correction techniques can be implemented to perform computation in the presence of errors \cite{preskill2018quantum}.


\subsection{Control and Readout Electronics}

The operation of quantum computers relies on sophisticated electronics serving two vital roles - control and readout. Control electronics manipulate the quantum processor by sending signals that drive qubit operations. Readout electronics measure the tiny signals produced by the qubits and convert them into usable data. Together, these systems provide the interface between the classical and quantum realms.

For superconducting qubits specifically, control is achieved by applying microwave pulses matched to the frequency of each qubit. Generating these precise control pulses requires arbitrary waveform generators with sampling rates exceeding 1 GHz to synthesize the pulse envelopes \cite{lu2017universal}. I/Q (in-phase/quadrature) modulators then provide control over the amplitude, phase, and frequency of a microwave carrier tone. The pulses are combined with the carrier waves through mixers and then up-converted or down-converted to the exact qubit frequencies. Programmable attenuators and amplifiers precisely tune the power levels to ensure accurate qubit manipulation. Switches are used to route pulses to specific qubits as needed. By gating the pulses in time, the switches ensure that operations are only applied to the intended target qubit. Low-frequency control electronics also assist by tuning device parameters and qubit frequencies for optimal performance. Proper coordination of these components enables high fidelity gates operations  \cite{lucero2008high}.

Qubit readout starts with faint microwave signals carrying information about the qubit state. Cryogenic amplifiers provide the first stage of amplification, critically boosting signal levels above thermal noise while adding minimal noise themselves \cite{devoret2013superconducting}. Semiconductor or superconducting parametric amplifiers are often used. Precise timing circuitry triggers the readout process and coordinates the capture of signals. The amplified signal is transmitted outside the cryogenic system for further processing. Additional amplifiers boost the signal, which is then filtered, digitized by high-speed analog-to-digital converters, and processed by digital logic. Field-programmable gate arrays oversee the signal processing, error correction, and integration with control electronics and classical computing resources. Careful calibration ensures high-fidelity qubit readout.

\subsection{Qubits}

Superconducting quantum computers are composed of superconducting circuits known as qubits, which leverage Josephson junctions (JJ) typically made from thin-film layers of aluminum or niobium. These JJs consist of two superconductors separated by an insulating barrier, enabling Cooper pairs to tunnel between the superconductors. When the junctions are cooled below their critical temperature (below 20 mK) they exhibit macroscopic quantum behavior \cite{martinis2005decoherence}. The most common superconducting qubit designs are transmons, flux qubits, and charge qubits. Transmon qubits rely on the nonlinear inductance of Josephson junctions shunted by a large capacitor to make their frequencies insensitive to noise. Flux qubits are composed of a superconducting loop with one or more JJs, where the quantum state depends on the magnetic flux threading the loop. Charge qubits encode information in the charge state of Cooper pairs on a small superconducting island connected to a reservoir via a JJ. These superconducting qubits are fabricated using standard lithographic techniques and deposited in thin film layers on a chip substrate. When cooled to mK temperatures in a dilution refrigerator, they can exist in a quantum superposition and be coherently manipulated by microwave pulses \cite{devoret2013superconducting}.

\subsection{Dilution Refrigerator and Associated Controls}

To reach the required operating temperatures below 20 mK, the quantum processor is mounted inside a dilution refrigerator. The fridge contains additional wiring, amplifiers, attenuators, filters, and other components to control and interact with the device while minimizing noise \cite{neill2013fluctuations}. Key components include pulse tube coolers to precool, a mixing chamber where the chip is mounted, and pumps to continuously circulate superfluid helium isotopes (helium-3 and helium-4) to cool the system. Temperature sensors, such as Ruthenium Oxide and Cernox, actively monitor the environment within the fridge, ensuring optimal conditions. Any deviations in temperature are countered by specialized controllers and heaters that adjust based on sensor readings \cite{krinner2019engineering}. For experiments demanding specific magnetic field conditions, integrated superconducting magnets, with their own set of control electronics, regulate the strength and direction of the applied field. The system is equipped with mechanisms to counteract the vibrations which can affect the delicate quantum processes. 
Modern dilution refrigerators also boast software interfaces, facilitating remote monitoring and adjustment, providing researchers with the flexibility to manage operations even from a distance \cite{IBMBluefors}.

\vspace{-10pt}
\section{IPs in the Quantum Computing System}

\textbf{IPs in materials and fabrication:} Quantum computing is a multifaceted field with many layers, each possessing unique IP aspects. At the heart of quantum computers are the qubits themselves. While companies openly share information about using superconducting circuits or ion traps, the specific materials and composition of these qubits represent closely-guarded trade secrets. For example, IBM publishes research on using transmon qubits made from niobium and aluminum. However, the exact shape, dimensions, treatment, and fabrication of these devices involves proprietary techniques fine-tuned by IBM through years of development \cite{mohseni2017commercialize}. These subtle details can impact qubit performance and reliability, representing the core IP.

\textbf{IPs in control electronics:} Another vital area is qubit control, which relies on precisely tailored microwave pulses applied to each qubit. Companies invest significantly in developing speciality control hardware and custom techniques to manage frequency crowding and prevent crosstalk errors between adjacent qubits. For example, companies may tweak pulse shapes, use optimal control algorithms, or add deliberate frequency offsets to qubits. These are crucial innovations that providers protect through patents and secrecy.

\textbf{IPs in quantum processor:} The qubit connectivity represented by coupling maps is also strategic intellectual property. While providers openly share coupling maps showing how qubits are interconnected, these maps likely differ from the physical hardware itself \cite{Cyber_workshop}. Companies may intentionally disconnect or hide certain couplings by disabling them in software. This selective disclosure might be an IP strategy, protecting information that could give insights into a company's unique quantum design \cite{kop2022intellectual}.

\textbf{IPs in cooling infrastructure:} The operational environment, including the dilution refrigerator, might be customized by each provider to create performance advantages. Companies may modify refrigerators by adding unique wiring, filters, heat shields, and control electronics tailored to their hardware \cite{crunkleton1992cryogenic}. For instance, they may tweak the mixing chamber shape and materials. The exact configurations, components used, or tweaks made by quantum service providers to these refrigerators can be proprietary, giving them an edge in system stability and performance.

\textbf{IPs in software and firmware:} 
Quantum compilers often possess proprietary methods for optimizing quantum circuits. Additionally, the strategies used for allocating programs from queues to processors, such as fair share allocation \cite{itoko2020scheduling}, might also contain IPs. However, such specifics can vary depending on the company.

\section{Qubit Technologies \& Their Supply Chains}
\label{sec:technologies}


This section provides a survey of the leading qubit technologies, their implementation, and the key players in the industry. Table \ref{tab:industry_map} compiles essential information on the supply chain infrastructure—including hardware, software, services, peripheral equipment, and firmware—of leading U.S.-based entities operating across diverse qubit platforms. 
The table further elucidates the specific components employed within these infrastructures and their providers. In instances where specific details regarding a company's supply chain components or providers are not publicly accessible or disclosed, we have denoted these entries within the table using ``NA". 

\input{very_big_table}

\subsection{Superconducting Qubits}
Superconducting qubits are a leading technology in the field of quantum computing, employed by companies such as IBM, Google, Rigetti Computing, and Bleximo \cite{kjaergaard2020superconducting}. 

\textbf{IBM: } 
IBM's quantum processors, made of superconducting transmon qubits, are housed in Yorktown Heights, New York \cite{IBMQuantum}. Along with their popular quantum software Qiskit, \cite{Qiskit}, IBM collaborates with other software giants like Zapata Computing, Strangeworks, QxBranch \cite{IBMStartupCollab}. The dilution fridges are procured from suppliers like Bluefors in Finland and Leiden Cryogenics in the Netherlands \cite{IBMBluefors}. 

\textbf{Google: } 
Google's Quantum AI campus is located in California, where they design and develop quantum processors using superconducting qubits \cite{GoogleQuantumLab}. 
They rely on dilution fridges from Lake Shore Cryotronics situated in Ohio \cite{GoogleCryoController}. 

\textbf{Rigetti Computing: } Rigetti Computing designs and manufactures its quantum processors using superconducting qubits, at their Fab-1 facility in California \cite{RigettiWhatWeBuild}. The company's recent flagship 84-qubit quantum computer, the Ankaa-1 84-qubit, exemplifies their hardware innovation. These efforts are supported by their own software platform, Forest, including pyQuil, quilc, and QVM \cite{PyQuilDocs}. 


\textbf{Intel: }
Intel's exploration into quantum computing encompasses not only superconducting but also spin qubits as a promising alternative \cite{IntelQuantumComputing, Guerreschi_2020}. 
Recently, Intel has made strides with the launch of the ``Tunnel Falls" quantum processor \cite{IntelQuantumComputing}, housing 12 qubits, a significant step in developing hardware to potentially surpass rivals.

\textbf{Bleximo: }Berkeley-based startup Bleximo specializes in building quantum accelerators and hardware technology for various applications, including cryptography and machine learning \cite{BleximoWebsite, BleximoCryogenic}. 

\subsection{Trapped Ions}

Trapped ions (TI) offer a promising avenue for scalable quantum computing by leveraging the innate properties of ions as qubits. Ions are confined in vacuum chambers using electromagnetic fields, isolating them for manipulation with lasers or microwaves \cite{haffner2008quantum}. 
The key advantage of trapped ions is the Coulombic interaction between them, allowing for consistent quantum information transfer and processing, thus making multi-qubit operations attainable. 

\textbf{Honeywell: } Honeywell Quantum Solutions, based in Colorado, specializes in TI quantum computing, using ytterbium ions as qubits \cite{HoneywellQuantum}. They have created a 10-qubit quantum computer with plans to reach 100 qubits by 2024. Honeywell has also devised quantum software tools, including a compiler and error correction software. 


\textbf{IonQ: } IonQ, headquartered in Maryland, is at the forefront of TI technology in quantum computing \cite{IonQTechnology}. IonQ Forte, the world's first software-configurable quantum computer, utilizes ytterbium ions and advanced acousto-optic deflectors (AODs) for precision and adaptability. 


\textbf{Alpine Quantum Technology: }
Alpine Quantum technology captures single-charged atoms within vacuum chambers using electric fields, turning each atom into a qubit, which is then precisely manipulated with laser pulses \cite{AQTTechnology}. 


\subsection{Topological Qubits}
Topological qubits present an intriguing pathway in quantum computing, utilizing topological states of matter to encode and protect quantum information. Unlike traditional qubits, topological qubits leverage the braiding of exotic particles like anyons and Majorana fermions, rendering them naturally resilient to local noise and disturbances \cite{nayak2008non}. 
\\
\textbf{Microsoft: }
Microsoft's foray into topological quantum computing focuses on Majorana fermions, particles not yet definitively observed but promising for error resilience \cite{MicrosoftQuantumQuest}. Their Quantum Lab at the University of California, Santa Barbara, has developed specialized cryogenic systems and fabrication processes. 

\subsection{Photonics-Based Quantum Computing}
Photonics-based quantum computing centers on modulating individual photon states to execute quantum functionalities, encompassing phenomena like superposition and entanglement \cite{takeda2019toward}. Generated primarily by lasers, these photons traverse through an assembly of optical components like beamsplitters, phase shifters and detectors. 
Key players like Xanadu, ORCA Computing, PsiQuantum, TundraSystems Global, Quandela, and QuiX Quantum are actively driving research and development in this field \cite{QuantumCompanies2023}.

\textbf{XANADU AI: }
Xanadu AI, headquartered in Canada, has developed hardware, particularly the Borealis and X-Series, which utilize photon-based squeezed state qubits \cite{XanaduWebsite}. Xanadu also offers the PennyLane software development kit coupled with Strawberry Fields—a groundbreaking open-source library curated for quantum machine learning \cite{PennyLaneWebsite}.



\vspace{-0.15cm}
\subsection{Quantum Annealing}


Quantum annealing is a sophisticated computational approach to address optimization challenges \cite{kadowaki1998quantum}. Unlike classical methods that examine one solution at a time, quantum annealing, with the help of superposition, explores a multitude of potential solutions simultaneously, leading to swifter convergence to the most optimal one. 

\textbf{D Wave \& Fujitsu: } 
D-Wave Systems, based in British Columbia, has innovated multiple quantum annealing systems, with the D-Wave Advantage2 being their most advanced release. These systems, known as D-Wave Quantum Processing Units (QPUs), utilize superconducting circuits to execute quantum annealing operations. D-Wave also offers the Ocean software development kit. Leap™ quantum cloud service provides enhanced accessibility to their quantum computing resources.\cite{DWaveQuantumLearning}

Fujitsu has developed Digital Annealer \cite{FujitsuDigitalAnnealer} which is a classical system equipped with a unique processor designed to emulate quantum annealing behaviors.



\vspace{-5pt}

\section{Quantum Supply Chain Security}
\label{sec:security}

\subsection{Leakage of IP}
From Table \ref{tab:industry_map}, it is evident that quantum computing systems may involve many third parties. This could present a risk to the IPs embedded in various layers of the stack. For example, calibration service providers may have access to the control electronics and pulsing and error mitigation strategies. Such information could be valuable to competitors and may be subject to leakage. Similarly, dilution fridge provider may have access to the quantum processor and the physical coupling map. Users have access to the compiler and the backends via APIs (Application Programming Interface). Although not shown, one can reverse engineer various compilation policies from access to the compiler.     

\subsection{QoS degradation}

Quantum computers rely on sophisticated peripherals like control/readout electronics, and cryogenic systems. Faults in these components can disrupt operations and reduce uptime. For example, a faulty microwave pulse generator that controls qubit gates would halt operations until fixed, severely impacting the QoS. Unlike classical cloud infrastructure, where redundancies provide resilience, most quantum providers operate singular, specialized systems with limited backup. 
These are highly specialized and customized systems involving unique qubit chips, control electronics, and peripheral equipment tailored for that particular device. There are no swappable redundant quantum systems waiting as backup making quantum systems far more vulnerable to supply chain disruptions 
than classical computing. 
Furthermore, hardware supply chain issues such as recycled and tampered parts can cause deliberate faults and leak information.

\subsection{Other Possible Attack Scenarios}
 \textbf{Hardware Weaknesses and Sensitivities: } 
The intricate quantum supply chain provides multiple avenues for potential manipulation of hardware properties by adversaries like malicious service providers with internal access. Firstly, qubits' fidelity and coherence time depends heavily on precise ambient conditions like temperature, noise, and magnetic shielding, which could be maliciously perturbed. Introducing even small controlled disturbances to these parameters, by compromising any part of the supply chain (e.g., untrusted parts within dilution refrigerator) responsible for maintaining these conditions, could deteriorate qubit performance. This can also create detectable changes in timing or power side-channels that attackers could exploit to extract information
such as user circuits or even reconstruct the full algorithm \cite{xu2023exploration}. Furthermore, properties like qubit connectivity, gate errors, and crosstalk are characterized during hardware calibration and testing before deployment. This data on error rates and qubit behavior is used when compiling and mapping quantum circuits onto the physical qubits. Maliciously altering such data can cause the compiler to make suboptimal or incorrect mapping decisions, assigning logical qubits to physical qubits differently 
\cite{acharya2020lightweight}.

\textbf{Peripheral Infiltration: }
The peripherals like dilution refrigerators and control/readout electronics in the quantum stack also present potential attack surfaces. For instance, the extreme cooling and isolation provided by fridges are critical for superconducting qubits \cite{krinner2019engineering}. Manipulation of fridge wiring or components by a compromised/malicious supplier could degrade qubit performance. Even minor changes to thermal regulation or noise filtering could kick qubits out of their fragile operating zones and open side-channels. Additionally, control electronics that generate microwave pulses rely on precise timing and synchronization. Small deviations by exploiting peripherals like arbitrary waveform generators, switches, or attenuators could corrupt pulse shapes. This can fail operations and create detectable anomalies for attackers to exploit. Further, any backdoors or kill switches introduced at the peripheral level could disable systems or wreck havoc at critical moments. The readout electronics are also susceptible. They amplify and process qubit measurements, which could be altered by compromising cryogenic amplifiers or ADC converters. Readout manipulation would corrupt program outputs. Sabotaged interconnects between peripherals could also disconnect vital components like amplifiers or magnets needed for qubit control.

\textbf{Firmware Threats: } 
The myriad firmware across the quantum stack also poses risks if compromised. For instance, control electronics use FPGAs running low-level firmware to coordinate qubit manipulations. Inserting backdoors could allow attackers to corrupt/manipulate pulse generation. Subtle perturbations to pulse parameters through firmware bugs can degrade fidelity enough to create side-channels. Refrigerator firmware that regulates qubit environment could similarly be targeted. Bugs inducing even small temperature or vibration changes may push qubits out of their operating sweet spot. Qubit calibration firmware also holds value and could be a target. Altering calibration could cripple performance or aid side-channels. 


\textbf{Software Vulnerabilities: } 
The intricate software stack supporting quantum computers presents multiple potential attack surfaces if compromised. Quantum algorithms initially expressed in high-level languages are compiled down to low-level pulse sequences executed on hardware. Inserting vulnerabilities during compilation could corrupt pulse parameters or scheduling in hard-to-detect ways. For example, a compromised compiler could manipulate gate decomposition or qubit mapping to degraded fidelity while maintaining functional correctness \cite{ghosh2023primer}. In addition to the compiler, software is responsible for orchestrating execution of user programs on quantum systems. This includes queuing submitted programs and allocating available quantum resources between users. The algorithms governing this scheduling and allocation could be manipulated to give certain users preferential treatment. For example, a backdoor in the scheduling software could allow specific users to jump ahead/behind in the queue or receive more/less processing time on quantum hardware. Therefore, some users could monopolize the system while denying services to others. Since quantum computer time is extremely limited, even small advantages in scheduling can be impactful. Furthermore, flaws in the design of APIs used to interface with quantum cloud platforms may exist. Attackers who discover these flaws could exploit the APIs to extract sensitive data about internal architectures or gain elevated privileges. Common API vulnerabilities like buffer overflows, command injections, or lack of input validation could allow adversaries to crash systems, manipulate outputs, or exfiltrate information. Additionally, quantum access is often provided via cloud-based services. Insufficient encryption and access controls on these cloud platforms further increase the attack risk. Attackers may intercept data, steal credentials, or gain persistence in cloud systems housing quantum software infrastructure \cite{trochatos2023hardware}.


\subsection{Defense Strategies}

Safeguarding the quantum supply chain requires multilayered defenses and controls spanning people, processes, components and technology. 
For personnel (including external third party service providers), the usual screening, background checks and contractual terms can prohibit leakage of proprietary information. Multi-party computation protocols may allow collaborative development without full IP disclosure. 
Process-wise, supplying vendors with only the minimal necessary data can reduce exposure. Regular audits can validate that proper controls and procedures are in place. Technology protections include obfuscation, access restrictions, data encryption, and watermarking IP to track dissemination. Formal verification of compiled code or firmware and enforcement of best code writing practices can protect from vulnerabilities that can enable code injection.

To minimize disruptions from faulty components and improve QoS, rigorous screening of parts and stress testing can be done. Redundant spares will allow rapid swap-in to restore services. Fail-safe designs may prevent single points of failure, and live redundancy techniques can provide constant backup systems. For quick recovery, effective repair management and supply processes are needed to enable prompt part replacement. Obfuscation to mask power or timing can address potential side channel leakages. 
To ensure reliable quantum computing with an unreliable or untrusted supply chain, one can continuously monitor and verify the integrity of critical components during operation. Additionally, using checksums, challenge-response protocols, and anomaly detection can help identify compromised components in real-time.

\vspace{-0.2cm}
\section{Conclusions}
\label{sec:conclusions}
We presented a first order analysis of quantum computing supply chain covering a wide range of qubit technologies. After reviewing the architecture of a representative quantum computing system, we presented various potential IPs embedded in various layers of the quantum stack. We also uncovered several potential security issues that have striking similarities with classical computing such as leakage of IP, QoS degradation and peripheral/software/firmware level attacks and a direction for defenses. Our study highlights the need to scrutinize the supply chain of quantum computers further to uncover potential vulnerabilities and threats and to build defenses proactively. 

\begin{acks}
The work is supported in parts by the National Science Foundation (NSF) (CNS-1722557, CCF-1718474, OIA-2040667, DGE-1723687 and DGE-1821766).
\end{acks}

\bibliographystyle{ACM-Reference-Format}
\bibliography{sample-base}

\appendix









\end{document}

%% file: very_big_table.tex
\begin{table*}[]
\vspace{-50pt}
\centering
\caption{The Quantum Industry Map}
\vspace{-10pt}

\begin{tabular}{c||c||c|c|c}
\hline
\textbf{\begin{tabular}[c]{@{}c@{}}Qubit\\ Technology\end{tabular}}                                             & \textbf{Company}                                                                      & \textbf{Infrastructure} & \textbf{Components}                                                                                 & \textbf{Provider}                                                                     \\ \hline \hline
\multirow{25}{*}{\textbf{\begin{tabular}[c]{@{}c@{}}Superconducting\\ Qubits\end{tabular}}}                     & \multirow{5}{*}{\textbf{IBM}}                                                         & Hardware                & Qubits/Quantum processors                                                                           & IBM Research Center                                                                   \\ \cline{3-5} 
                                                                                                                &                                                                                       & Software                & Compiler/Qiskit                                                                                     & IBM; other partners.                                                                  \\ \cline{3-5} 
                                                                                                                &                                                                                       & Services                & Calibration                                                                                         & IBM Quantum Support                                                                   \\ \cline{3-5} 
                                                                                                                &                                                                                       & Peripheral              & Dilution fridge                                                                                     & Bluefors and Leiden Cryogenics                                                        \\ \cline{3-5} 
                                                                                                                &                                                                                       & Firmware                & Control electronics                                                                                 & IBM                                                                                   \\ \cline{2-5} 
                                                                                                                & \multirow{5}{*}{\textbf{Google}}                                                      & Hardware                & Qubits/Quantum processors                                                                           & Google Research                                                                       \\ \cline{3-5} 
                                                                                                                &                                                                                       & Software                & QC Software Framework/Cirq                                                                          & Google                                                                                \\ \cline{3-5} 
                                                                                                                &                                                                                       & Services                & Quantum Cloud Services                                                                              & Google Cloud Platform                                                                 \\ \cline{3-5} 
                                                                                                                &                                                                                       & Peripheral              & Dilution fridge                                                                                     & Lake Shore Cryotronics                                                                \\ \cline{3-5} 
                                                                                                                &                                                                                       & Firmware                & NA                                                                                                  & NA                                                                                    \\ \cline{2-5} 
                                                                                                                & \multirow{5}{*}{\textbf{\begin{tabular}[c]{@{}c@{}}Rigetti\\ Computing\end{tabular}}} & Hardware                & Qubits/Quantum processors                                                                           & Rigetti Computing                                                                     \\ \cline{3-5} 
                                                                                                                &                                                                                       & Software                & Quantum Computing Software                                                                          & Rigetti Computing                                                                     \\ \cline{3-5} 
                                                                                                                &                                                                                       & Services                & Quantum Cloud Services                                                                              & Rigetti Quantum Cloud Services                                                        \\ \cline{3-5} 
                                                                                                                &                                                                                       & Peripheral              & Dilution fridge                                                                                     & Bluefors                                                                              \\ \cline{3-5} 
                                                                                                                &                                                                                       & Firmware                & NA                                                                                                  & NA                                                                                    \\ \cline{2-5} 
                                                                                                                & \multirow{5}{*}{\textbf{Intel}}                                                       & Hardware                & Superconducting \& Spin qubits                                                                      & \multirow{2}{*}{Intel}                                                                \\ \cline{3-4}
                                                                                                                &                                                                                       & Software                & Intel Quantum Simulator                                                                             &                                                                                       \\ \cline{3-5} 
                                                                                                                &                                                                                       & Services                & NA                                                                                                  & NA                                                                                    \\ \cline{3-5} 
                                                                                                                &                                                                                       & Peripheral              & \begin{tabular}[c]{@{}c@{}}Dilution fridge, Cryogenic Control \\ Systems (Horse Ridge)\end{tabular} & Bluefors, Intel, QuTech                                                               \\ \cline{3-5} 
                                                                                                                &                                                                                       & Firmware                & NA                                                                                                  & NA                                                                                    \\ \cline{2-5} 
                                                                                                                & \multirow{5}{*}{\textbf{Bleximo}}                                                     & Hardware                & Quantum accelerators                                                                                & \multirow{3}{*}{Bleximo}                                                              \\ \cline{3-4}
                                                                                                                &                                                                                       & Software                & Quantum Computing Software                                                                          &                                                                                       \\ \cline{3-4}
                                                                                                                &                                                                                       & Services                & Calibration/Training/Support                                                                        &                                                                                       \\ \cline{3-5} 
                                                                                                                &                                                                                       & Peripheral              & Dilution fridge                                                                                     & Bluefors, Oxford Instruments                                                          \\ \cline{3-5} 
                                                                                                                &                                                                                       & Firmware                & Control stack                                                                                       & Bleximo                                                                               \\ \hline \hline
\multirow{10}{*}{\textbf{Trapped Ions}}                                                                         & \multirow{5}{*}{\textbf{Honeywell}}                                                   & Hardware                & Qubits/Quantum processors                                                                           & \multirow{4}{*}{Honeywell Quantum Solutions}                                          \\ \cline{3-4}
                                                                                                                &                                                                                       & Software                & Quantum software tools                                                                              &                                                                                       \\ \cline{3-4}
                                                                                                                &                                                                                       & Services                & Access, consulting \& collaboration                                                                 &                                                                                       \\ \cline{3-4}
                                                                                                                &                                                                                       & Peripheral              & Cryogenic equipment                                                                                 &                                                                                       \\ \cline{3-5} 
                                                                                                                &                                                                                       & Firmware                & NA                                                                                                  & NA                                                                                    \\ \cline{2-5} 
                                                                                                                & \multirow{5}{*}{\textbf{IonQ}}                                                        & Hardware                & Qubits/Quantum processors                                                                           & IonQ                                                                                  \\ \cline{3-5} 
                                                                                                                &                                                                                       & Software                & Compiler/IonQ Cloud                                                                                 & IonQ; other partners                                                                  \\ \cline{3-5} 
                                                                                                                &                                                                                       & Services                & Calibration/Support                                                                                 & IonQ Quantum Support                                                                  \\ \cline{3-5} 
                                                                                                                &                                                                                       & Peripheral              & \begin{tabular}[c]{@{}c@{}}Cryogenic control electronics\\ and vacuum hardware\end{tabular}         & IonQ                                                                                  \\ \cline{3-5} 
                                                                                                                &                                                                                       & Firmware                & NA                                                                                                  & NA                                                                                    \\ \hline \hline
\multirow{5}{*}{\textbf{\begin{tabular}[c]{@{}c@{}}Topological\\ Qubits\end{tabular}}}                          & \multirow{5}{*}{\textbf{Microsoft}}                                                   & Hardware                & Topological qubits                                                                                  & Microsoft’s QC Lab                                                                    \\ \cline{3-5} 
                                                                                                                &                                                                                       & Software                & Q\# language, QDK                                                                                   & \begin{tabular}[c]{@{}c@{}}Quantum Architectures and\\ Computation Group\end{tabular} \\ \cline{3-5} 
                                                                                                                &                                                                                       & Services                & \begin{tabular}[c]{@{}c@{}}Optimization, enterprise-grade\\ security, hybrid cloud\end{tabular}     & \begin{tabular}[c]{@{}c@{}}Microsoft Azure\\ Quantum\end{tabular}                     \\ \cline{3-5} 
                                                                                                                &                                                                                       & Peripheral              & Dilution refrigerators                                                                              & Microsoft’s QC Lab                                                                    \\ \cline{3-5} 
                                                                                                                &                                                                                       & Firmware                & NA                                                                                                  & NA                                                                                    \\ \hline \hline
\multirow{5}{*}{\textbf{\begin{tabular}[c]{@{}c@{}}Photonics Based\\ Quantum\\ Computing\end{tabular}}}         & \multirow{5}{*}{\textbf{\begin{tabular}[c]{@{}c@{}}XANADU\\ AI\end{tabular}}}         & Hardware                & Borealis, X-Series                                                                                  & \multirow{3}{*}{Xanadu}                                                               \\ \cline{3-4}
                                                                                                                &                                                                                       & Software                & PennyLane, Strawberry Fields                                                                        &                                                                                       \\ \cline{3-4}
                                                                                                                &                                                                                       & Services                & Lightning, Jet simulators, Cloud                                                                    &                                                                                       \\ \cline{3-5} 
                                                                                                                &                                                                                       & Peripheral              & \multirow{2}{*}{NA}                                                                                 & \multirow{2}{*}{NA}                                                                   \\ \cline{3-3}
                                                                                                                &                                                                                       & Firmware                &                                                                                                     &                                                                                       \\ \hline \hline
\multirow{5}{*}{\textbf{\begin{tabular}[c]{@{}c@{}}Quantum Annealing\\ Based Quantum\\ Computing\end{tabular}}} & \multirow{5}{*}{\textbf{D-Wave}}                                                      & Hardware                & Qubits/Quantum processors                                                                           & \multirow{3}{*}{D-Wave Systems}                                                       \\ \cline{3-4}
                                                                                                                &                                                                                       & Software                & Ocean software development kit                                                                      &                                                                                       \\ \cline{3-4}
                                                                                                                &                                                                                       & Services                & Leap quantum cloud service                                                                          &                                                                                       \\ \cline{3-5} 
                                                                                                                &                                                                                       & Peripheral              & Cryogenic equipments                                                                                & Various suppliers                                                                     \\ \cline{3-5} 
                                                                                                                &                                                                                       & Firmware                & NA                                                                                                  & NA                                                                                    \\ \hline \hline
\end{tabular}

\label{tab:industry_map}
\end{table*}

%% file: sample-sigplan.bbl

\begin{thebibliography}{47}


\ifx \showCODEN    \undefined \def \showCODEN     #1{\unskip}     \fi
\ifx \showDOI      \undefined \def \showDOI       #1{#1}\fi
\ifx \showISBNx    \undefined \def \showISBNx     #1{\unskip}     \fi
\ifx \showISBNxiii \undefined \def \showISBNxiii  #1{\unskip}     \fi
\ifx \showISSN     \undefined \def \showISSN      #1{\unskip}     \fi
\ifx \showLCCN     \undefined \def \showLCCN      #1{\unskip}     \fi
\ifx \shownote     \undefined \def \shownote      #1{#1}          \fi
\ifx \showarticletitle \undefined \def \showarticletitle #1{#1}   \fi
\ifx \showURL      \undefined \def \showURL       {\relax}        \fi
\providecommand\bibfield[2]{#2}
\providecommand\bibinfo[2]{#2}
\providecommand\natexlab[1]{#1}
\providecommand\showeprint[2][]{arXiv:#2}

\bibitem[Acharya and Saeed(2020)]%
        {acharya2020lightweight}
\bibfield{author}{\bibinfo{person}{Nikita Acharya} {and}
  \bibinfo{person}{Samah~Mohamed Saeed}.} \bibinfo{year}{2020}\natexlab{}.
\newblock \showarticletitle{A lightweight approach to detect
  malicious/unexpected changes in the error rates of NISQ computers}. In
  \bibinfo{booktitle}{\emph{Proceedings of the 39th International Conference on
  Computer-Aided Design}}. \bibinfo{pages}{1--9}.
\newblock


\bibitem[Alberts et~al\mbox{.}(2021)]%
        {alberts2021accelerating}
\bibfield{author}{\bibinfo{person}{Garrelt~JN Alberts},
  \bibinfo{person}{M~Adriaan Rol}, \bibinfo{person}{Thorsten Last},
  \bibinfo{person}{Benno~W Broer}, \bibinfo{person}{Cornelis~C Bultink},
  \bibinfo{person}{Matthijs~SC Rijlaarsdam}, {and} \bibinfo{person}{Amber~E
  Van~Hauwermeiren}.} \bibinfo{year}{2021}\natexlab{}.
\newblock \showarticletitle{Accelerating quantum computer developments}.
\newblock \bibinfo{journal}{\emph{EPJ Quantum Technology}} \bibinfo{volume}{8},
  \bibinfo{number}{1} (\bibinfo{year}{2021}), \bibinfo{pages}{18}.
\newblock


\bibitem[AQT(2023)]%
        {AQTTechnology}
AQT \bibinfo{year}{2023}\natexlab{}.
\newblock \bibinfo{booktitle}{\emph{{AQT Technology}}}.
\newblock
\urldef\tempurl%
\url{https://www.aqt.eu/technology/}
\showURL{%
Retrieved August 13, 2023 from \tempurl}


\bibitem[Bleximo(2022)]%
        {BleximoCryogenic}
\bibfield{author}{\bibinfo{person}{Bleximo}.} \bibinfo{year}{2022}\natexlab{}.
\newblock \bibinfo{booktitle}{\emph{{Cryogenic Hardware at Bleximo:
  Superconducting Quantum Processor Packaging and Shielding}}}.
\newblock
\urldef\tempurl%
\url{https://medium.com/bleximo/cryogenic-hardware-at-bleximo-superconducting-quantum-processor-packaging-and-shielding-d0d2b84b47d3}
\showURL{%
Retrieved August 13, 2023 from \tempurl}


\bibitem[Bleximo(2023)]%
        {BleximoWebsite}
Bleximo \bibinfo{year}{2023}\natexlab{}.
\newblock \bibinfo{booktitle}{\emph{{Powering Innovation Through Quantum
  Computing}}}.
\newblock
\urldef\tempurl%
\url{https://bleximo.com/}
\showURL{%
Retrieved August 13, 2023 from \tempurl}


\bibitem[Cross et~al\mbox{.}(2017)]%
        {cross2017open}
\bibfield{author}{\bibinfo{person}{Andrew~W Cross}, \bibinfo{person}{Lev~S
  Bishop}, \bibinfo{person}{John~A Smolin}, {and} \bibinfo{person}{Jay~M
  Gambetta}.} \bibinfo{year}{2017}\natexlab{}.
\newblock \showarticletitle{Open quantum assembly language}.
\newblock \bibinfo{journal}{\emph{arXiv preprint arXiv:1707.03429}}
  (\bibinfo{year}{2017}).
\newblock


\bibitem[Crunkleton(1992)]%
        {crunkleton1992cryogenic}
\bibfield{author}{\bibinfo{person}{James~A Crunkleton}.}
  \bibinfo{year}{1992}\natexlab{}.
\newblock \bibinfo{booktitle}{\emph{Cryogenic refrigeration apparatus}}.
\newblock \bibinfo{type}{{T}echnical {R}eport}. \bibinfo{institution}{Boreas
  Inc}.
\newblock


\bibitem[Dargan(2022)]%
        {QuantumCompanies2023}
\bibfield{author}{\bibinfo{person}{James Dargan}.}
  \bibinfo{year}{2022}\natexlab{}.
\newblock \bibinfo{booktitle}{\emph{{81 Quantum Computing Companies: An
  Ultimate 2023 List}}}.
\newblock
\urldef\tempurl%
\url{https://thequantuminsider.com/2022/09/05/quantum-computing-companies-ultimate-list-for-2022/}
\showURL{%
Retrieved August 13, 2023 from \tempurl}


\bibitem[Devoret and Schoelkopf(2013)]%
        {devoret2013superconducting}
\bibfield{author}{\bibinfo{person}{Michel~H Devoret} {and}
  \bibinfo{person}{Robert~J Schoelkopf}.} \bibinfo{year}{2013}\natexlab{}.
\newblock \showarticletitle{Superconducting circuits for quantum information:
  an outlook}.
\newblock \bibinfo{journal}{\emph{Science}} \bibinfo{volume}{339},
  \bibinfo{number}{6124} (\bibinfo{year}{2013}), \bibinfo{pages}{1169--1174}.
\newblock


\bibitem[DWave(2023)]%
        {DWaveQuantumLearning}
DWave \bibinfo{year}{2023}\natexlab{}.
\newblock \bibinfo{booktitle}{\emph{{How D-Wave Systems Work }}}.
\newblock
\urldef\tempurl%
\url{https://www.dwavesys.com/learn/quantum-computing/}
\showURL{%
Retrieved August 13, 2023 from \tempurl}


\bibitem[Fujitsu(2023)]%
        {FujitsuDigitalAnnealer}
Fujitsu \bibinfo{year}{2023}\natexlab{}.
\newblock \bibinfo{booktitle}{\emph{{Fujitsu Digital Annealer}}}.
\newblock
\urldef\tempurl%
\url{https://www.fujitsu.com/global/services/business-services/digital-annealer/}
\showURL{%
Retrieved August 13, 2023 from \tempurl}


\bibitem[Ghosh et~al\mbox{.}(2023)]%
        {ghosh2023primer}
\bibfield{author}{\bibinfo{person}{Swaroop Ghosh}, \bibinfo{person}{Suryansh
  Upadhyay}, {and} \bibinfo{person}{Abdullah~Ash Saki}.}
  \bibinfo{year}{2023}\natexlab{}.
\newblock \showarticletitle{A Primer on Security of Quantum Computing}.
\newblock \bibinfo{journal}{\emph{arXiv preprint arXiv:2305.02505}}
  (\bibinfo{year}{2023}).
\newblock


\bibitem[GoogleQuantumLab(2023)]%
        {GoogleQuantumLab}
GoogleQuantumLab \bibinfo{year}{2023}\natexlab{}.
\newblock \bibinfo{booktitle}{\emph{{Google Quantum AI: Our Lab}}}.
\newblock
\urldef\tempurl%
\url{https://quantumai.google/hardware/our-lab}
\showURL{%
Retrieved August 13, 2023 from \tempurl}


\bibitem[Guerreschi et~al\mbox{.}(2020)]%
        {Guerreschi_2020}
\bibfield{author}{\bibinfo{person}{Gian~Giacomo Guerreschi},
  \bibinfo{person}{Justin Hogaboam}, \bibinfo{person}{Fabio Baruffa}, {and}
  \bibinfo{person}{Nicolas P~D Sawaya}.} \bibinfo{year}{2020}\natexlab{}.
\newblock \showarticletitle{Intel Quantum Simulator: a cloud-ready
  high-performance simulator of quantum circuits}.
\newblock \bibinfo{journal}{\emph{Quantum Science and Technology}}
  \bibinfo{volume}{5}, \bibinfo{number}{3} (\bibinfo{date}{may}
  \bibinfo{year}{2020}), \bibinfo{pages}{034007}.
\newblock
\urldef\tempurl%
\url{https://doi.org/10.1088/2058-9565/ab8505}
\showDOI{\tempurl}


\bibitem[H{\"a}ffner et~al\mbox{.}(2008)]%
        {haffner2008quantum}
\bibfield{author}{\bibinfo{person}{Hartmut H{\"a}ffner},
  \bibinfo{person}{Christian~F Roos}, {and} \bibinfo{person}{Rainer Blatt}.}
  \bibinfo{year}{2008}\natexlab{}.
\newblock \showarticletitle{Quantum computing with trapped ions}.
\newblock \bibinfo{journal}{\emph{Physics reports}} \bibinfo{volume}{469},
  \bibinfo{number}{4} (\bibinfo{year}{2008}), \bibinfo{pages}{155--203}.
\newblock


\bibitem[Honeywell(2023)]%
        {HoneywellQuantum}
Honeywell \bibinfo{year}{2023}\natexlab{}.
\newblock \bibinfo{booktitle}{\emph{{Honeywell Quantum}}}.
\newblock
\urldef\tempurl%
\url{https://www.honeywell.com/us/en/company/quantum}
\showURL{%
Retrieved August 13, 2023 from \tempurl}


\bibitem[IBM Quantum(2023)]%
        {IBMQuantum}
IBM Quantum \bibinfo{year}{2023}\natexlab{}.
\newblock \bibinfo{booktitle}{\emph{{IBM Quantum Computing Research}}}.
\newblock
\urldef\tempurl%
\url{https://research.ibm.com/quantum-computing}
\showURL{%
Retrieved August 13, 2023 from \tempurl}


\bibitem[Intel(2023)]%
        {IntelQuantumComputing}
Intel \bibinfo{year}{2023}\natexlab{}.
\newblock \bibinfo{booktitle}{\emph{{Intel Quantum Computing}}}.
\newblock
\urldef\tempurl%
\url{https://www.intel.com/content/www/us/en/research/quantum-computing.html}
\showURL{%
Retrieved August 13, 2023 from \tempurl}


\bibitem[IonQ(2023)]%
        {IonQTechnology}
IonQ \bibinfo{year}{2023}\natexlab{}.
\newblock \bibinfo{booktitle}{\emph{{IonQ Technology}}}.
\newblock
\urldef\tempurl%
\url{https://ionq.com/technology}
\showURL{%
Retrieved August 13, 2023 from \tempurl}


\bibitem[Itoko and Imamichi(2020)]%
        {itoko2020scheduling}
\bibfield{author}{\bibinfo{person}{Toshinari Itoko} {and}
  \bibinfo{person}{Takashi Imamichi}.} \bibinfo{year}{2020}\natexlab{}.
\newblock \showarticletitle{Scheduling of operations in quantum compiler}. In
  \bibinfo{booktitle}{\emph{2020 IEEE International Conference on Quantum
  Computing and Engineering (QCE)}}. IEEE, \bibinfo{pages}{337--344}.
\newblock


\bibitem[Kadowaki and Nishimori(1998)]%
        {kadowaki1998quantum}
\bibfield{author}{\bibinfo{person}{Tadashi Kadowaki} {and}
  \bibinfo{person}{Hidetoshi Nishimori}.} \bibinfo{year}{1998}\natexlab{}.
\newblock \showarticletitle{Quantum annealing in the transverse Ising model}.
\newblock \bibinfo{journal}{\emph{Physical Review E}} \bibinfo{volume}{58},
  \bibinfo{number}{5} (\bibinfo{year}{1998}), \bibinfo{pages}{5355}.
\newblock


\bibitem[Kjaergaard et~al\mbox{.}(2020)]%
        {kjaergaard2020superconducting}
\bibfield{author}{\bibinfo{person}{Morten Kjaergaard},
  \bibinfo{person}{Mollie~E Schwartz}, \bibinfo{person}{Jochen Braum{\"u}ller},
  \bibinfo{person}{Philip Krantz}, \bibinfo{person}{Joel I-J Wang},
  \bibinfo{person}{Simon Gustavsson}, {and} \bibinfo{person}{William~D
  Oliver}.} \bibinfo{year}{2020}\natexlab{}.
\newblock \showarticletitle{Superconducting qubits: Current state of play}.
\newblock \bibinfo{journal}{\emph{Annual Review of Condensed Matter Physics}}
  \bibinfo{volume}{11} (\bibinfo{year}{2020}), \bibinfo{pages}{369--395}.
\newblock


\bibitem[Kop et~al\mbox{.}(2022)]%
        {kop2022intellectual}
\bibfield{author}{\bibinfo{person}{Mauritz Kop}, \bibinfo{person}{Mateo Aboy},
  {and} \bibinfo{person}{Timo Minssen}.} \bibinfo{year}{2022}\natexlab{}.
\newblock \showarticletitle{Intellectual property in quantum computing and
  market power: a theoretical discussion and empirical analysis}.
\newblock \bibinfo{journal}{\emph{Journal of Intellectual Property Law and
  Practice}} \bibinfo{volume}{17}, \bibinfo{number}{8} (\bibinfo{year}{2022}),
  \bibinfo{pages}{613--628}.
\newblock


\bibitem[Krinner et~al\mbox{.}(2019)]%
        {krinner2019engineering}
\bibfield{author}{\bibinfo{person}{Sebastian Krinner}, \bibinfo{person}{Simon
  Storz}, \bibinfo{person}{Philipp Kurpiers}, \bibinfo{person}{Paul Magnard},
  \bibinfo{person}{Johannes Heinsoo}, \bibinfo{person}{Raphael Keller},
  \bibinfo{person}{Janis Luetolf}, \bibinfo{person}{Christopher Eichler}, {and}
  \bibinfo{person}{Andreas Wallraff}.} \bibinfo{year}{2019}\natexlab{}.
\newblock \showarticletitle{Engineering cryogenic setups for 100-qubit scale
  superconducting circuit systems}.
\newblock \bibinfo{journal}{\emph{EPJ Quantum Technology}} \bibinfo{volume}{6},
  \bibinfo{number}{1} (\bibinfo{year}{2019}), \bibinfo{pages}{2}.
\newblock


\bibitem[Lu et~al\mbox{.}(2017)]%
        {lu2017universal}
\bibfield{author}{\bibinfo{person}{Yao Lu}, \bibinfo{person}{Srivatsan
  Chakram}, \bibinfo{person}{Ngainam Leung}, \bibinfo{person}{Nathan Earnest},
  \bibinfo{person}{Ravi~K Naik}, \bibinfo{person}{Ziwen Huang},
  \bibinfo{person}{Peter Groszkowski}, \bibinfo{person}{Eliot Kapit},
  \bibinfo{person}{Jens Koch}, {and} \bibinfo{person}{David~I Schuster}.}
  \bibinfo{year}{2017}\natexlab{}.
\newblock \showarticletitle{Universal stabilization of a parametrically coupled
  qubit}.
\newblock \bibinfo{journal}{\emph{Physical review letters}}
  \bibinfo{volume}{119}, \bibinfo{number}{15} (\bibinfo{year}{2017}),
  \bibinfo{pages}{150502}.
\newblock


\bibitem[Lucero et~al\mbox{.}(2008)]%
        {lucero2008high}
\bibfield{author}{\bibinfo{person}{Erik Lucero}, \bibinfo{person}{Max
  Hofheinz}, \bibinfo{person}{Markus Ansmann}, \bibinfo{person}{Radoslaw~C
  Bialczak}, \bibinfo{person}{N Katz}, \bibinfo{person}{Matthew Neeley},
  \bibinfo{person}{AD O’Connell}, \bibinfo{person}{H Wang},
  \bibinfo{person}{AN Cleland}, {and} \bibinfo{person}{John~M Martinis}.}
  \bibinfo{year}{2008}\natexlab{}.
\newblock \showarticletitle{High-fidelity gates in a single Josephson qubit}.
\newblock \bibinfo{journal}{\emph{Physical review letters}}
  \bibinfo{volume}{100}, \bibinfo{number}{24} (\bibinfo{year}{2008}),
  \bibinfo{pages}{247001}.
\newblock


\bibitem[Martinis et~al\mbox{.}(2005)]%
        {martinis2005decoherence}
\bibfield{author}{\bibinfo{person}{John~M Martinis}, \bibinfo{person}{Ken~B
  Cooper}, \bibinfo{person}{Robert McDermott}, \bibinfo{person}{Matthias
  Steffen}, \bibinfo{person}{Markus Ansmann}, \bibinfo{person}{KD Osborn},
  \bibinfo{person}{Katarina Cicak}, \bibinfo{person}{Seongshik Oh},
  \bibinfo{person}{David~P Pappas}, \bibinfo{person}{Raymond~W Simmonds},
  {et~al\mbox{.}}} \bibinfo{year}{2005}\natexlab{}.
\newblock \showarticletitle{Decoherence in Josephson qubits from dielectric
  loss}.
\newblock \bibinfo{journal}{\emph{Physical review letters}}
  \bibinfo{volume}{95}, \bibinfo{number}{21} (\bibinfo{year}{2005}),
  \bibinfo{pages}{210503}.
\newblock


\bibitem[Microsoft(2015)]%
        {MicrosoftQuantumQuest}
Microsoft \bibinfo{year}{2015}\natexlab{}.
\newblock \bibinfo{booktitle}{\emph{{The Quantum Quest at Microsoft}}}.
\newblock
\urldef\tempurl%
\url{https://www.microsoft.com/en-us/research/blog/the-quantum-quest-at-microsoft/}
\showURL{%
Retrieved August 13, 2023 from \tempurl}


\bibitem[Mohseni et~al\mbox{.}(2017)]%
        {mohseni2017commercialize}
\bibfield{author}{\bibinfo{person}{Masoud Mohseni}, \bibinfo{person}{Peter
  Read}, \bibinfo{person}{Hartmut Neven}, \bibinfo{person}{Sergio Boixo},
  \bibinfo{person}{Vasil Denchev}, \bibinfo{person}{Ryan Babbush},
  \bibinfo{person}{Austin Fowler}, \bibinfo{person}{Vadim Smelyanskiy}, {and}
  \bibinfo{person}{John Martinis}.} \bibinfo{year}{2017}\natexlab{}.
\newblock \showarticletitle{Commercialize quantum technologies in five years}.
\newblock \bibinfo{journal}{\emph{Nature}} \bibinfo{volume}{543},
  \bibinfo{number}{7644} (\bibinfo{year}{2017}), \bibinfo{pages}{171--174}.
\newblock


\bibitem[Nayak et~al\mbox{.}(2008)]%
        {nayak2008non}
\bibfield{author}{\bibinfo{person}{Chetan Nayak}, \bibinfo{person}{Steven~H
  Simon}, \bibinfo{person}{Ady Stern}, \bibinfo{person}{Michael Freedman},
  {and} \bibinfo{person}{Sankar~Das Sarma}.} \bibinfo{year}{2008}\natexlab{}.
\newblock \showarticletitle{Non-Abelian anyons and topological quantum
  computation}.
\newblock \bibinfo{journal}{\emph{Reviews of Modern Physics}}
  \bibinfo{volume}{80}, \bibinfo{number}{3} (\bibinfo{year}{2008}),
  \bibinfo{pages}{1083}.
\newblock


\bibitem[Neill et~al\mbox{.}(2013)]%
        {neill2013fluctuations}
\bibfield{author}{\bibinfo{person}{C Neill}, \bibinfo{person}{A Megrant},
  \bibinfo{person}{R Barends}, \bibinfo{person}{Yu Chen}, \bibinfo{person}{B
  Chiaro}, \bibinfo{person}{J Kelly}, \bibinfo{person}{JY Mutus},
  \bibinfo{person}{PJJ O'Malley}, \bibinfo{person}{D Sank}, \bibinfo{person}{J
  Wenner}, {et~al\mbox{.}}} \bibinfo{year}{2013}\natexlab{}.
\newblock \showarticletitle{Fluctuations from edge defects in superconducting
  resonators}.
\newblock \bibinfo{journal}{\emph{Applied Physics Letters}}
  \bibinfo{volume}{103}, \bibinfo{number}{7} (\bibinfo{year}{2013}).
\newblock


\bibitem[Nielsen and Chuang(2001)]%
        {nielsen2001quantum}
\bibfield{author}{\bibinfo{person}{Michael~A Nielsen} {and}
  \bibinfo{person}{Isaac~L Chuang}.} \bibinfo{year}{2001}\natexlab{}.
\newblock \showarticletitle{Quantum computation and quantum information}.
\newblock \bibinfo{journal}{\emph{Phys. Today}} \bibinfo{volume}{54},
  \bibinfo{number}{2} (\bibinfo{year}{2001}), \bibinfo{pages}{60}.
\newblock


\bibitem[O'Shea(2021)]%
        {IBMBluefors}
\bibfield{author}{\bibinfo{person}{Dan O'Shea}.}
  \bibinfo{year}{2021}\natexlab{}.
\newblock \bibinfo{booktitle}{\emph{{IBM-Bluefors partnership promises really
  cool quantum future}}}.
\newblock
\urldef\tempurl%
\url{https://www.insidequantumtechnology.com/news-archive/ibm-bluefors-partnership-promises-really-cool-quantum-future/}
\showURL{%
Retrieved August 13, 2023 from \tempurl}


\bibitem[PennyLane Website(2023)]%
        {PennyLaneWebsite}
PennyLane Website \bibinfo{year}{2023}\natexlab{}.
\newblock \bibinfo{booktitle}{\emph{{PennyLane}}}.
\newblock
\urldef\tempurl%
\url{https://pennylane.ai/}
\showURL{%
Retrieved August 13, 2023 from \tempurl}


\bibitem[Pittsburg Quantum Institute(2023)]%
        {Cyber_workshop}
Pittsburg Quantum Institute \bibinfo{year}{2023}\natexlab{}.
\newblock \bibinfo{booktitle}{\emph{{Workshop on Cybersecurity of Quantum
  Computing}}}.
\newblock
\urldef\tempurl%
\url{https://www.pqi.org/news/workshop-cybersecurity-quantum-computing}
\showURL{%
Retrieved August 13, 2023 from \tempurl}


\bibitem[Preskill(2018)]%
        {preskill2018quantum}
\bibfield{author}{\bibinfo{person}{John Preskill}.}
  \bibinfo{year}{2018}\natexlab{}.
\newblock \showarticletitle{Quantum computing in the NISQ era and beyond}.
\newblock \bibinfo{journal}{\emph{Quantum}}  \bibinfo{volume}{2}
  (\bibinfo{year}{2018}), \bibinfo{pages}{79}.
\newblock


\bibitem[PyQuil(2023)]%
        {PyQuilDocs}
PyQuil \bibinfo{year}{2023}\natexlab{}.
\newblock \bibinfo{booktitle}{\emph{{PyQuil Documentation}}}.
\newblock
\urldef\tempurl%
\url{https://pyquil-docs.rigetti.com/en/v2.7.2/}
\showURL{%
Retrieved August 13, 2023 from \tempurl}


\bibitem[Qiskit(2023)]%
        {Qiskit}
Qiskit \bibinfo{year}{2023}\natexlab{}.
\newblock \bibinfo{booktitle}{\emph{{Qiskit}}}.
\newblock
\urldef\tempurl%
\url{https://qiskit.org/}
\showURL{%
Retrieved August 13, 2023 from \tempurl}


\bibitem[Rigetti(2023)]%
        {RigettiWhatWeBuild}
Rigetti \bibinfo{year}{2023}\natexlab{}.
\newblock \bibinfo{booktitle}{\emph{{Rigetti: What We Build}}}.
\newblock
\urldef\tempurl%
\url{https://www.rigetti.com/what-we-build}
\showURL{%
Retrieved August 13, 2023 from \tempurl}


\bibitem[Rol(2020)]%
        {rol2020control}
\bibfield{author}{\bibinfo{person}{Michiel~Adriaan Rol}.}
  \bibinfo{year}{2020}\natexlab{}.
\newblock \showarticletitle{Control for programmable superconducting quantum
  systems}.
\newblock  (\bibinfo{year}{2020}).
\newblock


\bibitem[Rostami et~al\mbox{.}(2014)]%
        {rostami2014primer}
\bibfield{author}{\bibinfo{person}{Masoud Rostami}, \bibinfo{person}{Farinaz
  Koushanfar}, {and} \bibinfo{person}{Ramesh Karri}.}
  \bibinfo{year}{2014}\natexlab{}.
\newblock \showarticletitle{A primer on hardware security: Models, methods, and
  metrics}.
\newblock \bibinfo{journal}{\emph{Proc. IEEE}} \bibinfo{volume}{102},
  \bibinfo{number}{8} (\bibinfo{year}{2014}), \bibinfo{pages}{1283--1295}.
\newblock


\bibitem[Takeda and Furusawa(2019)]%
        {takeda2019toward}
\bibfield{author}{\bibinfo{person}{Shuntaro Takeda} {and}
  \bibinfo{person}{Akira Furusawa}.} \bibinfo{year}{2019}\natexlab{}.
\newblock \showarticletitle{Toward large-scale fault-tolerant universal
  photonic quantum computing}.
\newblock \bibinfo{journal}{\emph{APL Photonics}} \bibinfo{volume}{4},
  \bibinfo{number}{6} (\bibinfo{year}{2019}).
\newblock


\bibitem[Trochatos et~al\mbox{.}(2023)]%
        {trochatos2023hardware}
\bibfield{author}{\bibinfo{person}{Theodoros Trochatos},
  \bibinfo{person}{Chuanqi Xu}, \bibinfo{person}{Sanjay Deshpande},
  \bibinfo{person}{Yao Lu}, \bibinfo{person}{Yongshan Ding}, {and}
  \bibinfo{person}{Jakub Szefer}.} \bibinfo{year}{2023}\natexlab{}.
\newblock \showarticletitle{Hardware Architecture for a Quantum Computer
  Trusted Execution Environment}.
\newblock \bibinfo{journal}{\emph{arXiv preprint arXiv:2308.03897}}
  (\bibinfo{year}{2023}).
\newblock


\bibitem[Wesler(2018)]%
        {IBMStartupCollab}
\bibfield{author}{\bibinfo{person}{Jeff Wesler}.}
  \bibinfo{year}{2018}\natexlab{}.
\newblock \bibinfo{booktitle}{\emph{{IBM Collaborating With Top Startups to
  Accelerate Quantum Computing}}}.
\newblock
\urldef\tempurl%
\url{https://newsroom.ibm.com/IBM-research?item=30420}
\showURL{%
Retrieved August 13, 2023 from \tempurl}


\bibitem[Wiggers(2019)]%
        {GoogleCryoController}
\bibfield{author}{\bibinfo{person}{Kyle Wiggers}.}
  \bibinfo{year}{2019}\natexlab{}.
\newblock \bibinfo{booktitle}{\emph{{Google’s new cryogenic quantum
  controller uses less than 2 milliwatts}}}.
\newblock
\urldef\tempurl%
\url{https://venturebeat.com/mobile/googles-new-cryogenic-quantum-controller-uses-less-than-2-milliwatts/}
\showURL{%
Retrieved August 13, 2023 from \tempurl}


\bibitem[Xanadu Website(2023)]%
        {XanaduWebsite}
Xanadu Website \bibinfo{year}{2023}\natexlab{}.
\newblock \bibinfo{booktitle}{\emph{{Xanadu}}}.
\newblock
\urldef\tempurl%
\url{https://www.xanadu.ai/}
\showURL{%
Retrieved August 13, 2023 from \tempurl}


\bibitem[Xu et~al\mbox{.}(2023)]%
        {xu2023exploration}
\bibfield{author}{\bibinfo{person}{Chuanqi Xu}, \bibinfo{person}{Ferhat Erata},
  {and} \bibinfo{person}{Jakub Szefer}.} \bibinfo{year}{2023}\natexlab{}.
\newblock \showarticletitle{Exploration of Quantum Computer Power
  Side-Channels}.
\newblock \bibinfo{journal}{\emph{arXiv preprint arXiv:2304.03315}}
  (\bibinfo{year}{2023}).
\newblock


\end{thebibliography}
